\documentclass{tlp}
\usepackage{aopmath}
\usepackage{amstext}
\usepackage{url}
\usepackage{times}
\usepackage{enumerate}
\usepackage{paralist}
\usepackage{ifpdf}
\ifpdf
\usepackage{microtype}
\fi

\title[Catching the Ouroboros: On Debugging Non-ground Answer-Set Programs]{Catching the Ouroboros: On Debugging Non-ground Answer-Set Programs}
\author[J. Oetsch, J. P\"uhrer, and H. Tompits]{JOHANNES OETSCH, J\"ORG P\"UHRER, AND HANS TOMPITS%
\thanks{This work was partially supported by the Austrian Science Fund~(FWF) under grant P21698.} \\ 
Technische Universit\"at Wien,\\
Institut f\"ur Informationssysteme~184/3,\\
Favoritenstrasse 9-11, A-1040 Vienna,
Austria,\\
\email{\{oetsch,puehrer,tompits\}@kr.tuwien.ac.at}}

\newcommand{\meta}{\Gamma}
\newcommand{\reif}[1]{\Delta(#1)}
\renewcommand{\L}{\mathcal{L}}

\newcommand{\commadots}{,\ldots,}
\newcommand{\constant}[1]{\ensuremath{\mathit{#1}}}
\newcommand{\naf}{\mathrm{not}}
\newcommand{\posbody}{B^{+}}
\newcommand{\negbody}{B^{-}}
\newcommand{\head}{H}
\newcommand{\grd}{\mathit{ground}}
\renewcommand{\grd}{\mathit{grd}}
\newcommand{\subst}{\vartheta}

\newcommand{\wrt}{with respect to}
\newcommand{\iec}{i.e.,\ }
\newcommand{\egc}{e.g.,\ }

\newcommand{\NP}{\mbox{NP}}
\newcommand{\coNP}{\mbox{co-NP}}
\newcommand{\D}{\mbox{D}}

\newcommand{\nop}[1]{}          %

\newcommand{\reifsymb}{\varrho}          %
\newcommand{\reifrule}[1]{\reifsymb_\mathit{rule}(#1)}          %
\newcommand{\reifprg}[1]{\reifsymb_\mathit{prg}(#1)}          %
\newcommand{\reifint}[1]{\reifsymb_\mathit{int}(#1)}          %

\newcommand{\INPUT}{\reif{P,I}}          %

\newcommand{\Unsat}{\mathit{UNSAT}} %
\newcommand{\UnsatGuess}{\mathit{UNSAT_{guess}}} %
\newcommand{\UnsatCheck}{\mathit{UNSAT_{check}}} %
\newcommand{\UnsatAux}{\mathit{UNSAT_{aux}}} %
\renewcommand{\Unsat}{\gamma_\mathit{unsat}} %
\renewcommand{\UnsatGuess}{\Unsat^\mathit{guess}} %
\renewcommand{\UnsatCheck}{\Unsat^\mathit{check}} %
\renewcommand{\UnsatAux}{\Unsat^\mathit{aux}} %

\newcommand{\Loop}{\mathit{LOOP}} %
\newcommand{\LoopGuess}{\mathit{LOOP_{guess}}} %
\newcommand{\LoopCheck}{\mathit{LOOP_{check}}} %
\newcommand{\LoopAux}{\mathit{LOOP_{aux}}} %
\renewcommand{\Loop}{\gamma_\mathit{loop}} %
\renewcommand{\LoopGuess}{\Loop^\mathit{guess}} %
\renewcommand{\LoopCheck}{\Loop^\mathit{check}} %
\renewcommand{\LoopAux}{\Loop^\mathit{aux}} %

\newcommand{\Support}{\mathit{SUPPORT}} %
\newcommand{\SupportGuess}{\mathit{SUPPORT_{guess}}} %
\newcommand{\SupportCheck}{\mathit{SUPPORT_{check}}} %
\newcommand{\SupportAux}{\mathit{SUPPORT_{aux}}} %
\renewcommand{\Support}{\gamma_\mathit{unfd}} %
\renewcommand{\SupportGuess}{\Support^\mathit{guess}} %
\renewcommand{\SupportCheck}{\Support^\mathit{check}} %
\renewcommand{\SupportAux}{\Support^\mathit{aux}} %

\newcommand{\Cons}{\mathit{CONS}}
\renewcommand{\Cons}{\gamma_\mathit{cons}}

\renewcommand{\mathit}[1]{{\text{\it{#1}}}}

\newtheorem{definition}{Definition}

\begin{document}
\label{firstpage}

\submitted{7 February 2010}
\revised{{\rm (}n/a{\rm )}}
\accepted{20 March 2010}

\maketitle


\begin{abstract}
An important issue towards a broader acceptance of answer-set programming (ASP) is the deployment of tools which support the programmer during the coding phase. In particular, methods for \emph{debugging} an answer-set program are recognised as a crucial step in this regard. Initial work on debugging in ASP mainly focused on propositional programs, yet practical debuggers need to handle programs with variables as well. In this paper, we discuss a debugging technique that is directly geared towards non-ground programs. Following previous work, we address the central debugging question why some interpretation is not an answer set. The explanations provided by our method are computed by means of a meta-programming technique, using a uniform encoding of a debugging request in terms of ASP itself. Our method also permits programs containing comparison predicates and integer arithmetics, thus covering a relevant language class commonly supported by all state-of-the-art ASP solvers.
 \end{abstract}

\begin{keywords}
answer-set programming, program analysis, debugging
\end{keywords}

\section{Introduction}

During the last decade, answer-set programming (ASP) has become a well-acknowledged paradigm for 
declarative problem solving. 
Although there exist efficient solvers (see, \egc \citeANP{competition09}~\citeyear{competition09} for an overview) and  a considerable body of literature 
concerning the theoretical foundations of ASP,
 comparably little effort has been spent
on methods to support the development of ASP programs.
Especially novice  programmers, tempted by the intuitive semantics and expressive power of 
ASP, may get disappointed and discouraged soon when
some observed program behaviour diverges from his or her expectations. 
Unlike for other programming languages like Java or C++, there is currently little support for \emph{debugging} a program
 in ASP, \iec  methods to \emph{explain} and \emph{localise} unexpected observations. This  is a clear  shortcoming of ASP and work in this direction has already started~\cite{brain05,syrjaenen06,brain07,mirek07,CaballeroGS08,gebser08,pontelli09,WittocxVD09}.

Most of the current debugging approaches for ASP rely on declarative strategies, focusing on \emph{conceptual errors} of programs, \iec mismatches between the intended meaning and the actual meaning of a program.
In fact, an elegant realisation of declarative debugging is to use ASP itself to debug  programs in ASP. This has been put forth, \egc in the approaches of \citeANP{brain07}~\citeyear{brain07} and \citeANP{gebser08}~\citeyear{gebser08}.
While the former uses a ``tagging'' method to decompose a program and applying various debugging queries, the latter is based on a meta-programming technique, \iec using a program over a meta-language to manipulate a program over an object language (in this case, both the meta-language and the object language are instances of ASP).
Such techniques have the obvious benefits of allowing (i)~to use reliable state-of-the-art ASP solvers as back-end reasoning engines and (ii)~to stay within the same paradigm for both the programming and debugging process.
Indeed, both approaches are realised by the system \texttt{spock}~\cite{spock07}.
However, like most other ASP debugging proposals, \texttt{spock} can deal only with propositional programs which is clearly a limiting factor as far as practical applications are concerned.

In this paper, we present a debugging method for non-ground programs following the methodology of the meta-programming approach of \citeANP{gebser08}~\citeyear{gebser08} for propositional programs.
That is to say, we deal with the problem of finding reasons why some interpretation is \emph{not} an answer set of a given program.
This is addressed by referring to a model-theoretic
characterisation of answer sets due to \citeANP{lee05} \citeyear{lee05}:
An interpretation $I$ is not an answer set of a 
program $P$ iff
(i)~some rule in $P$ is not classically satisfied by $I$ or
(ii)~$I$ contains some loop of $P$ that is unfounded by  $P$ \wrt\ $I$.
Intuitively,  Item~(ii) states that some atoms in $I$ are not justified by $P$ in the  sense that
no rules in $P$ can derive them or that some atoms are in $I$ only because they are derived by a set of rules in a  circular way---like the \emph{Ouroboros}, the ancient symbol of a dragon biting its own tail that represents cyclicality and eternity.
This characterisation seems to be quite natural and intuitive for \emph{explaining}
why some interpretation is not an answer set. 
Furthermore, a particular 
benefit is that it can ease the subsequent  \emph{localisation} of errors
since the witnesses why an interpretation is not an answer set, like
rules which are not satisfied, unfounded atoms,  or cyclic rules  responsible for unfounded loops,
can be located in the program or the interpretation.

Although, at first glance, one may be inclined to directly apply the original approach of \citeANP{gebser08}~\citeyear{gebser08}
to programs with variables by simply
grounding them in a preprocessing step, one problem in such an endeavour is that then it is not immediate clear how to 
relate explanations for the propositional program to the non-ground program. The more severe problem, however,
is that the grounding step requires exponential space and time \wrt\ the size of the problem instance which yields a mismatch of the overall complexity as
checking whether an interpretation is an answer set of some (non-ground) program is complete
for $\Pi^{P}_{2}$~\cite{eiter04}, and thus the complementary problem why some interpretation is not
an answer set is complete for $\Sigma^{P}_{2}$---our method to decide this problem accounts
for this complexity bound and avoids exponential space requirements.
Indeed, we devise a \emph{uniform} encoding of our basic debugging problem in terms of a \emph{fixed} disjunctive logic program $\meta$ and an efficient reification of 
a problem instance as a set $\reif{P,I}$ of facts, where $P$ is the program to be debugged and $I$ is the interpretation under consideration.
Explanations why $I$ is not an answer set of $P$ are then obtained by the answer sets of $\meta\cup \reif{P,I}$.

We stress that the definition of  $\meta$ is  non-trivial: while the meta-program in the approach of \citeANP{gebser08}~\citeyear{gebser08} for debugging propositional disjunctive programs
could be achieved in terms of a normal non-ground program, \emph{by uniformly encoding a $\Sigma^{P}_{2}$ property, we reach the very limits of disjunctive ASP} and have to rely on advanced saturation techniques that inherently require disjunctions in rule heads~\cite{eiter97}.

Currently, our approach handles disjunctive logic programs with  constraints,
integer arithmetic, comparison predicates, and strong negation, thus covering 
a practically relevant program class. Further language constructs, in particular
aggregates and weak constraints, are left for future work.

\section{Preliminaries}\label{sec:prel}

We deal with \emph{disjunctive logic programs}  which are finite sets of rules of form
\[
a_1 \vee \cdots \vee a_l \leftarrow a_{l+1}, \ldots, a_{m}, \naf\ a_{m+1}, \ldots, \naf\ a_{n} , 
\]
where $n \geq m \geq l \geq 0$, ``$\naf$'' denotes \emph{default negation}, and  all $a_i$ are literals over 
a function-free first-order language $\L$. A literal is an atom possibly preceded by the \emph{strong negation} symbol $\neg$.
In the sequel, we assume that $\L$ will be implicitly defined by the considered programs.
For a rule $r$ as above, we define the \emph{head}  of $r$ as $\head(r) = \{a_1,  \ldots,  a_l\}$, 
the \emph{positive body} as $\posbody(r) = \{a_{l+1}, \ldots, a_{m}\}$, and  the \emph{negative body} as $\negbody(r) = \{a_{m+1}, \ldots, a_{n}\}$. 
If $n=l=1$,
$r$  is a \emph{fact}; if $r$ contains no disjunction, $r$ is \emph{normal}; and
if $l=0$ and $n>0$,
$r$ is a \emph{constraint}. For facts, we will omit the symbol $\leftarrow$.
A literal, rule, or program is \emph{ground} if it contains no variables.
Furthermore, a program is normal if all rules in it are normal. 
Finally, we allow arithmetic and comparison predicate symbols $+$, $*$, $=$, $\neq$, $\leq$, $<$, $\geq$, and~$>$ in programs, but these may appear only positively in rule bodies.

Let $C$ be a set of constants. A \emph{substitution over} $C$ is a function $\subst$ assigning each variable an element of $C$.
We denote by $e\subst$ the result of applying $\subst$ to an expression $e$.
The \emph{grounding} of a  program $P$ relative to its Herbrand universe, denoted by $\grd(P)$, is defined as usual.

An \emph{interpretation} $I$ (over some language $\L$) is a finite and consistent set of ground literals (over $\L$) that does not contain any arithmetic or comparison predicates.
Recall that consistency means that $\{a,\neg a\}\not\subseteq I$, for any atom $a$.
The satisfaction relation, $I\models \alpha$, between $I$ and a ground atom, a literal, 
a rule, a set of literals, or a program $\alpha$ is defined in the usual manner.
Note that the presence of arithmetic and comparison operators implies that the domain of our language will normally include natural numbers as well as a linear ordering, $\preceq$, 
for evaluating the comparison relations (which coincides with the usual ordering in case of constants which are natural numbers).

For any ground program $P$ and any interpretation $I$, the \emph{reduct}, $P^{I}$, of $P$ \wrt\ $I$ \cite{gelfond91}
is defined as 
$P^{I} = \{\head(r) \leftarrow \posbody(r) \mid r \in P, I \cap \negbody(r) = \emptyset \}$.
An interpretation $I$ is an \emph{answer set} of a program $P$ iff $I$ is  a minimal model of $\grd(P)$.

We will base our subsequent elaboration on an alternative characterisation of answer sets following \citeN{lee05}, described next.
Given a program $P$,  the \emph{positive dependency graph} is a directed
graph $(V,A)$, where 
(i)~$V$ equals the Herbrand base of the considered language $\L$ and
(ii)~$(a,b) \in A$ iff $a \in \head(r)$ and $b \in \posbody(r)$, for some rule $r 
\in \grd(P)$.
A non-empty set $L$ of ground literals is  a \emph{loop}\footnote{Note that loops have first been studied by \citeANP{lin04}~\citeyear{lin04}; different definitions of loops for non-ground programs were given by \citeANP{chen06}~\citeyear{chen06} and \citeANP{lee08}~\citeyear{lee08}.
For our purposes, it suffices to refer to the basic definition for ground programs.} of  a  program $P$ iff,
for each pair $a,b \in L$,
there is a path $\pi$ of length greater than or equal to 0 from $a$ to $b$
in the positive dependency graph of $P$ such that each literal in $\pi$
is in $L$. 

Let $P$ be a program and $I$ and $J$ interpretations. Then,
$J$ is \emph{externally supported by $P$ \wrt\ $I$} iff
there is a rule $r \in \grd(P)$ 
such that
(i)~$I\models\posbody(r)$ and $I\cap \negbody(r)=\emptyset$,
(ii)~$\head(r) \cap J \neq \emptyset$,
(iii)~$(\head(r)\setminus J)\cap I=\emptyset$, and
(iv)~$\posbody(r) \cap J = \emptyset$.

Intuitively, Items (i)--(iii) express that $J$ is supported by $P$ \wrt\ $I$, in the sense that the grounding of $P$ contains some rule $r$ whose body is satisfied by $I$ (Item~(i)) and which is able to derive some literal in $J$ (Item~(ii)), while all head atoms of $r$ not contained in $J$ are false under $I$.
Moreover, Item~(iv) ensures that this support is external as it is without
reference to the set $J$ itself.

Answer sets are now characterised thus:

\begin{proposition}[\citeNPS{lee05}]\label{prop:lee05}
Let $P$ be a program and $I$ an interpretation. Then,
$I$ is an answer set of $P$ iff
{\rm (}i{\rm )}~$I \models P$ and
{\rm (}ii{\rm )}~every loop of $P$ that is contained in $I$ is externally supported by $P$ \wrt~$I$.
\end{proposition}

We actually make mainly use of the complementary relation of external support: Following \citeN{LeoneRS97}, we call $J$ \emph{unfounded by $P$ \wrt\ $I$} iff $J$ is not externally supported by 
$P$ \wrt\ $I$. 

\section{The basic debugging approach}

As discussed in the introduction, 
we view  an error as a mismatch between the intended answer sets and the observed actual answer sets of some program. 
More specifically, our basic debugging question is why a given interpretation $I$ is not answer set of some program $P$, and
thus we deal with  finding explanations for $I$ not being an answer set of $P$.
Proposition~\ref{prop:lee05} allows us to distinguish between two kinds of such explanations:
(i)~instantiations of rules in $P$ that are not satisfied by $I$ and (ii)~loops of $P$ in $I$ that are unfounded by $P$ \wrt\ $I$.
Although our basic debugging question
allows for different, multi-faceted, answers,
we see two major benefits of referring to this kind of  categorisation:
First, in view of Proposition~\ref{prop:lee05}, these kinds of explanations are
always sufficient to explain why $I$ is not an answer set of $P$, and second,
this method provides \emph{concrete witnesses}, \egc unsatisfied rules or unfounded atoms,
that can help to localise the reason for an error in a program or an interpretation in a rather intuitive way.

Before we introduce the details of our approach, 
we discuss its virtues compared to a method for debugging non-ground programs which can be obtained using the previous meta-programming technique for propositional programs due to \citeN{gebser08}.

\subsection{Prelude: A case for directly debugging non-ground programs}

Explaining why some interpretation is not an answer set of some program based on the characterisation of \citeN{lee05} has been dealt with in previous work for 
debugging propositional disjunctive logic programs~\cite{gebser08}. 
In principle, we could use this method for debugging non-ground programs as well by employing a preparatory  grounding step.
However, such an undertaking comes at a higher computational cost compared to our approach which respects the
inherent complexity of the underlying tasks.
We lay down our arguments in what follows.

To begin with, let us recall that \citeN{gebser08}
defined a fixed normal non-ground program $\gamma$  and a mapping $\delta$ from disjunctive propositional programs  and interpretations 
to sets of facts. 
Given a disjunctive program $P$ without variables and some interpretation $I$, explanations
why $I$ is not an answer set of $P$ can then be extracted from the answer sets of $\gamma \cup \delta(P,I)$. 
Such a problem encoding is 
\emph{uniform} in the sense that $\gamma$ does not depend on the problem instance determined by $P$ and $I$.

To find reasons why some interpretation $I$ is not an answer set of a non-ground program $P$, 
the above approach can be used by computing the answer sets of $\gamma \cup \delta(\grd(P),I)$.
However, in general, the size of $\grd(P)$ is exponential in the size of $P$, and the computation 
of the answer sets of a ground program requires exponential time with respect
to the size of the program, unless the polynomial hierarchy collapses.
Hence, this outlined approach to compute explanations using a grounding step requires, all in all, \emph{exponential space} and \emph{double-exponential time} with respect to the size of $P$.
But this is a mismatch to the inherent complexity of the overall task, as the following result shows: 

\begin{proposition}
\label{prop:complexity}
Given a program $P$ and an interpretation $I$, deciding whether $I$ is not an answer set of $P$ is $\Pi^P_2$-complete. 
\end{proposition}
This property is a consequence of the well-known fact that the complementary problem, \iec checking whether some given interpretation is an answer set of some program, is $\Sigma^{P}_{2}$-complete~\cite{eiter04}.
Hence, checking whether an interpretation is not an answer set of some program 
can be computed in \emph{polynomial space}.

Our approach takes this complexity property into account.  We exploit the expressive power of disjunctive non-ground ASP by providing a
uniform encoding that avoids both exponential space and double-exponential time requirements:
Given a program $P$ and an interpretation $I$, we define an encoding $\meta \cup \reif{P,I}$, where $\meta$ is a fixed disjunctive non-ground program, and $\reif{P,I}$ is an efficient encoding  of $P$ and $I$ by means of facts.
Explanations why $I$ is not an answer set of $P$ are determined by the answer sets of $\meta \cup \reif{P,I}$.
Since $\meta$ is fixed, the grounding of $\meta \cup \reif{P,I}$ is bounded by a polynomial in the size of $P$ and $I$. Thus, our
approach requires only polynomial space and single-exponential time with respect to $P$ and $I$.

Note that disjunctions can presumably not be avoided in $\meta$ due to the  $\Pi^{P}_{2}$-hardness of 
deciding whether an interpretation is not an answer set of some program.
One may ask, however, whether $\meta$ could be normal in case $P$ is normal.
We have to answer in the negative: answer-set checking for normal programs is complete for $\D^{P}$, 
even if no negation is used or negation is only used in a stratified way~\cite{eiter04}.
(We recall that $\D^{P}$ is the class of problems that  can be  decided by a conjunction
of an $\NP$ and an independent  $\coNP$ property.)
Hence, $\meta$ cannot be normal unless $\NP = \coNP$.
However, one could use two independent normal meta-programs to encode our desired task.

A further benefit of debugging a program  directly at the non-ground level is that we can immediately 
relate explanations for errors to first-order expressions in  the considered program,
\egc to rules or literals with variables instead
of their ground instantiations.

In what follows, we give details of $\meta$ and $\Delta$ and describe their main properties.

\subsection{Construction of the meta-program}
\subsubsection{Reification of input instances}
For realising the encoding $\INPUT$ for program $P$ and interpretation $I$,
we rely on a reification $\reifprg{P}$ of $P$ and a reification $\reifint{I}$ of $I$.
The former is, in turn, constructed from reifications $\reifrule{r}$ of each individual rule $r\in P$.
We introduce the mappings $\reifrule{\cdot}$, $\reifprg{\cdot}$, and $\reifint{\cdot}$ in the following.

To begin with, we need unique names for certain language elements.
By an \emph{extended predicate symbol} (EPS) we understand a predicate symbol, possibly preceded by  
the symbol for strong negation.
Let 
$\cdot'$ be an injective \emph{labelling function} from the set of program rules, literals, EPSs, and variables to a set of labels
from the symbols in our language $\L$. 
Note that we do not need labels for constant symbols since they will serve as unique names for themselves.

A single program rule  is reified by means of facts according to the following definition.
\begin{definition}
Let  $r$ be a rule. Then,
\begin{center}

$\begin{array}{r@{}l}
\reifrule{r} =\ & \{ \mathit{rule}(r') \} \cup  \{ \mathit{head}(r', a') \mid a \in \head(r) \} \cup \\
&  \{ \mathit{posbody}(r', a') \mid a \in \posbody(r) \} \cup 
     \{ \mathit{negbody}(r', a') \mid a \in \negbody(r) \} \cup \\
 &   \{ \mathit{pred}(a',L') \mid \mbox{$a = L(x_{1},\ldots,x_{n})$} \mbox{ is a literal in $r$},  
    \mbox{$L$ is an EPS}\} \cup \\
 &  \{ \mathit{struct}(a',i,\constant{var},x_{i}') \mid \mbox{$a\!=\!L(x_{1},\ldots, x_{n})$}   \mbox{ is a literal in $r$,} 
     \mbox{ $L$ is an EPS,}\\  
  &  \hspace{1em} \mbox{$i \in \{1, \ldots, n\}$, and $x_{i}$ is a variable}  \} \cup \\ 
 & \{ \mathit{struct}(a',i,\constant{const},x_{i}) \mid   \mbox{$a = L(x_{1},\ldots, x_{n})$}  \mbox{ is a literal}    
   \mbox{ in $r$, $L$ is an EPS,}\\
  &\hspace{1em} \mbox{$i \in \{1, \ldots, n\}$, and $x_{i}$ is a constant symbol}  \} \cup \\
 & \{ \mathit{var}(r',x') \mid \mbox{$x$ is a variable occurring in $r$} \}\mbox{.}
\end{array}$
\end{center}
\end{definition}
The first fact states that label $r'$ denotes a rule.
The next three sets of facts associate labels of the literals in the head, the positive body, and the negative body
to the respective parts of~$r$.
Then, each label of some literal in $r$ is associated with a label for its  EPS.
The following two sets of facts encode the positions of variables and constants in the literals of the rule.
Finally, the last set of facts states which variables occur in the rule $r$.

A program is encoded as follows:
\begin{definition}
Let $P$ be a program. Then, 
\begin{center}
$\begin{array}{r@{}l}
\reifprg{P} =\ &
\bigcup_{r \in P} \reifrule{r} \cup 
\{ \mathit{dom}(c) \mid \mbox{$c$ is  a constant  symbol in $P$} \} \cup \\
& \{ \mathit{arity}(L',n) \mid \mbox{a = $L(x_1, \ldots, x_n)$ is a literal in $P$}, 
\mbox{$L$ is an EPS}  \}\mbox{.}
\end{array}$
\end{center}
\end{definition}

 The first union of facts stem from the reification of the single rules in the program.
The remaining facts  represent the Herbrand universe of the program and 
associate the EPSs occurring in the program with their
arities.

The translation from an interpretation to a set of facts is formalised by the next definition.
\begin{definition}
Let  $I$ be an interpretation. Then, 
\begin{center}
$\begin{array}{r@{}l}
\reifint{I} =\ & \{\mathit{int}(a') \mid  a \in I\}  \cup  
 \{\mathit{pred}(a',L') \mid \mbox{$a = L(x_{1}, \ldots, x_{n})$ is a literal in $I$,} \\
&\quad  \mbox{$L$ is an EPS}\}   \cup \\
&\{\mathit{struct}(a',i,\constant{const},x_{i}) \mid   \mbox{$a = L(x_{1},\ldots, x_{n})$ is a literal in $I$, $L$ is an EPS, }\\
&\quad \mbox{$i \in \{1, \ldots, n\}$, and $x_{i}$ is a constant symbol}  \}\mbox{.} 
\end{array}$
\end{center}
\end{definition}

The first two sets of facts associate the literals in  $I$ with their respective labels and 
EPSs.
The last set of facts reifies the internal structure of the literals occurring in $I$. 

\begin{definition}
Let $P$ be a program and $I$ an interpretation. 
Furthermore, let $N$ be the  the maximum of $|I|$ and the arities of all predicate symbols in $P$.
Then, $\INPUT = 
 \reifprg{P} \cup \reifint{I}  \cup 
\{ \mathit{natNumber}(n) \mid n \in \{0, \ldots, N\} \}$.

\end{definition}
The literals $\mathit{natNumber}(\cdot)$  are necessary  to add sufficiently many natural numbers to the Herbrand universe of $\INPUT$ to carry out correctly all computations in the subsequent program encodings. 
Note that the size of $\INPUT$  is always linear in the size
of $P$ and~$I$.

\begin{figure}[t]
\begin{small}
\hrule
$$\begin{array}{r@{}l}
\UnsatGuess  = \{ & \mathit{guessRule}(R) \vee \mathit{nguessRule}(R) \leftarrow \mathit{rule}(R), 
\\
&
\mathit{someRule} \leftarrow \mathit{guessRule}(R),\\
&
\leftarrow \naf\ \mathit{someRule}, \mathit{rule}(R),
\\
&
\leftarrow \mathit{guessRule}(R_{1}), \mathit{guessRule}(R_{2}), R_{1} \neq R_{2}, \\
 &\mathit{subst}(X,C) \vee \mathit{nsubst}(X,C) \leftarrow \mathit{guessRule}(R), \mathit{var}(R,X), \mathit{dom}(C), \\
 &\mathit{assigned}(X) \leftarrow \mathit{subst}(X,C),
\\
  & 
\leftarrow \naf\ \mathit{assigned}(X), \mathit{guessRule}(R),   \mathit{var}(R,X), \\
&\leftarrow  \mathit{subst}(X,C_{1}),  \mathit{subst}(X,C_{2}), C_{1} \neq C_{2} \}\mbox{.}\\[.3em]
\UnsatCheck  =  \{ &   \mathit{unsatisfied} \leftarrow  \mathit{satBody}, \naf\ \mathit{satHead},
\\
&
\mathit{satBody} \leftarrow  \naf\ \mathit{unsatPosbody}, \naf\ \mathit{unsatNegbody},\\
&\mathit{satHead}      \leftarrow \mathit{guessRule}(R), \mathit{head}(R,A), \mathit{true}(A),\\
&\mathit{unsatPosbody} \leftarrow \mathit{guessRule}(R),\mathit{posbody}(R,A), 
             \mathit{false}(A),\\
&\begin{array}{@{}r}
\mathit{unsatNegbody} \leftarrow \mathit{guessRule}(R),\mathit{negbody}(R,A), 
                          \mathit{true}(A) \}\mbox{.}
\end{array}
\end{array}$$\\
\hrule
\end{small}
\caption{Modules $\UnsatGuess$ and  $\UnsatCheck$\label{fig:UNSAT}.}
\end{figure}

\subsubsection{The meta-program $\meta$}

We proceed with the definition of the central meta-program $\Gamma$. The complete program 
consists of more than 160 rules. For space reasons, we only present the relevant parts and omit 
modules containing simple auxiliary definitions.
The full encodings can be found at 
\begin{quote}
\url{www.kr.tuwien.ac.at/research/projects/mmdasp/encoding.tar.gz}.
\end{quote}

The meta-program $\meta$ consists of the following modules:
(i)~$\Unsat$, related to unsatisfied rules,
(ii)~$\Loop$, related to loops,
(iii)~$\Support$, for testing unfoundedness of loops, and
(iv)~$\Cons$, integrating Parts~(i)--(iii) for performing the overall test of whether a given interpretation $I$ is not an answer set of 
a given program $P$.

We  first introduce the program module $\Unsat$ to identify unsatisfied rules. 

\begin{definition}\label{def:unsat}
By $\Unsat$ we understand the program 
$\UnsatGuess \cup \UnsatCheck \cup \UnsatAux$, where $\UnsatGuess$ and $\UnsatCheck$ are given in Figure~\ref{fig:UNSAT}, and $\UnsatAux$ defines the auxiliary predicates $\mathit{true}(\cdot)$ and $\mathit{false}(\cdot)$. 
\end{definition}

Intuitively, for a  program $P$ and an interpretation $I$, $\UnsatGuess$
guesses a rule $r \in P$, represented by predicate $\mathit{guessRule}$, 
and a substitution $\subst$, represented by $\mathit{subst}$, and
  $\UnsatCheck$ defines that
  $\mathit{unsatisfied}$ holds  if $I \not\models r\subst$. 
Module $\UnsatAux$ (omitted for space reasons) defines the auxiliary predicates $\mathit{true}(\cdot)$ and $\mathit{false}(\cdot)$ such that
$\mathit{true}(l')$ holds if $I \models l\subst$, for some literal $l$, and $\mathit{false}(l')$ holds if $I \not\models l\subst$.

Module $\Unsat$ has the following central property:

\begin{theorem}\label{thm:unsat}
Let $P$ be a program and $I$ an interpretation.
Then, $I \not\models P$ iff some answer set of $\Unsat \cup \INPUT$ contains $\mathit{unsatisfied}$.
More specifically,
for each rule  $r \in P$  with $I \not\models r\subst$, for some substitution $\subst$ over the Herbrand universe of $P$,
$\Unsat \cup \INPUT$ has an answer set $S$ such that
{\rm (}i{\rm )}~$\{\mathit{unsatisfied},\mathit{guessRule}(r')\}\subseteq S$ and
{\rm (}ii{\rm )}~$\mathit{subst}(x',c)\in S$ 
iff $\subst(x) = c$.
\end{theorem}

We next define  module $\Loop$ for identifying 
loops of a program.

\begin{definition}\label{def:loop}
By $\Loop$ we understand the program $ \LoopGuess  \cup \LoopCheck \cup \LoopAux$, where $\LoopGuess$ and $\LoopCheck$ are given in Figure~\ref{fig:LOOP}, and $\LoopAux$ defines the auxiliary predicates $\mathit{loopSz}(\cdot)$ and $\mathit{differSeq}(\cdot,\cdot,\cdot)$. 
%
%
\begin{figure}[t]
\begin{small}
\hrule
$$\begin{array}{r@{}l}
 \LoopGuess  =  \{ & \mathit{inLoop}(X) \vee \mathit{outLoop}(X) \leftarrow \mathit{int}(X),
\\
&
\mathit{someInLoop} \leftarrow \mathit{inLoop}(X),
\\
&
\leftarrow \naf\ \mathit{someInLoop}, \mathit{int}(X) \}\mbox{.}\\[.3em]
\LoopCheck  =  \{ &  
 \begin{array}[t]{@{}l@{}l}
 \mathit{inRuleSet}(N,R) \vee \mathit{outRuleSet}(N,R) \leftarrow\ & 1 \leq N, N \leq S,
		  \mathit{loopSz}(S), \mathit{rule}(R), \\&
		  \mathit{natNumber}(N),
\end{array}\\
& \mathit{someRule}(N) \leftarrow \mathit{inRuleSet}(N,R), \\
&\leftarrow \naf\ \mathit{someRule}(N), 1 \leq N, N \leq S, \mathit{loopSz}(S),  \mathit{rule}(R), \mathit{natNumber}(N), \\
&\leftarrow \mathit{inRuleSet}(N,R_{1}), \mathit{inRuleSet}(N,R_{2}), R_{1} \neq R_{2},  \\ 
&\leftarrow  \mathit{inRuleSet}(N_{1},R_{1}), \mathit{inRuleSet}(N_{2},R_{2}), N_{1} \leq N_{2}, R_{1} > R_{2}, \\
&\begin{array}[t]{@{}l@{}l}
\mathit{loopSubst}(N,X,C) \vee \mathit{nloopSubst}(N,X,C) \leftarrow\ &\mathit{var}(R,X),\mathit{dom}(C),\\&
\mathit{inRuleSet}(N,R),\end{array}\\
&\mathit{loopAssigned}(N,X) \leftarrow \mathit{loopSubst}(N,X,C), \\ 
&\leftarrow \naf\ \mathit{loopAssigned}(N,X), \mathit{inRuleSet}(N,R), \mathit{var}(R,X), \\ 
&\leftarrow \mathit{loopSubst}(N,X,C_{1}), \mathit{loopSubst}(N,X,C_{2}), C_{1} \neq C_{2},  \\
&\mathit{isLoop}  \leftarrow \naf\ \mathit{unreachablePair} , inLoop(X), \\ 
&\mathit{unreachablePair} \leftarrow \mathit{inLoop}(X), \mathit{inLoop}(Y),  \naf\ \mathit{path}(X,Y), \\ 
&\mathit{path}(X,X) \leftarrow \mathit{inLoop}(X),   \\ 
&\begin{array}[t]{@{}l@{}l}
\mathit{path}(X,Y) \leftarrow\ & \mathit{inLoop}(X), \mathit{inLoop}(Y), \mathit{pred}(X,T_{1}), \mathit{pred}(Y,T_{2}),
       	       \mathit{loopSz}(S), \\& 1 \leq N, N \leq S, 
	       \mathit{head}(R,H),\mathit{inRuleSet}(N,R), 
	         \mathit{posbody}(R,B), \\&\mathit{pred}(H,T_{1}), \mathit{pred}(B,T_{2}),  
	          \naf\ \mathit{differSeq}(N,X,H), \\& \naf\ \mathit{differSeq}(N,Y,B),
\end{array}\\
&\mathit{path}(X,Z) \leftarrow \mathit{inLoop}(X), \mathit{inLoop}(Z), \mathit{path}(X,Y), \mathit{path}(Y,Z) \}\mbox{.}
\end{array}$$
\hrule
\end{small}
\caption{Modules $\LoopGuess$ and  $\LoopCheck$\label{fig:LOOP}.}
\end{figure}
\end{definition}

Intuitively, for a program $P$ and an interpretation $I$,
$\LoopGuess$  guesses  a non-empty subset $L$ of $I$, represented by  $\mathit{inLoop}(\cdot)$, 
as  a candidate for a loop, and
$\LoopCheck$ defines that $\mathit{isLoop}$ holds if $L$ is a loop of $P$.
More specifically, this check is realised as follows.
Assume $L$ contains $n$ literals.

\begin{enumerate}
\item Guess a set $\mathcal{G}$ of $n$ pairs $(r, \subst)$, where $r$ is a rule from $P$ and $\subst$ is a substitution over the Herbrand universe of $P$.

\item Check, for each $a,b \in L$, whether there is a path $\pi$ 
in the positive dependency graph of the ground program consisting of rules 
$\{r\subst \mid (r, \subst) \in \mathcal{G}\}$ such that
$\pi$ starts with  $a$ and ends with $b$, and
all literals in $\pi$ are in $L$. 
A path $\pi$ is represented by the binary  predicate $\mathit{path}(\cdot,\cdot)$.
\end{enumerate}

Module $\LoopAux$ (again omitted for space reasons) defines that (i)~$\mathit{loopSz}(n)$ holds if $|L| = n$ and (ii)~$\mathit{differSeq}(i,a',b')$ holds  if $a\subst \neq b\subst$, where  $a$, $b$ are literals and
$\subst$ is the substitution stemming from a pair in $\mathcal{G}$ that is associated with an index $i$ by $\Loop$.

\begin{theorem}\label{thm:loop}
For any program $P$ and any interpretation $I$,
$L \subseteq I$ is a loop of $P$ iff,
for some answer set $S$ of $\Loop \cup \INPUT$, $\mathit{isLoop} \in S$ and $L = \{x \mid \mathit{inLoop}(x') \in S\}$.
\end{theorem}

We proceed with module $\Support$ for
checking whether some set  $J$
of ground literals  is unfounded by $P$ \wrt\ an interpretation $I$.  
We later combine this $\coNP$ check  with $\Loop$ to
identify unfounded loops, \iec we will integrate a loop guess with a $\coNP$ check, thus reaching the very limits of disjunctive ASP by uniformly
encoding a $\Sigma^{P}_{2}$ property.

\begin{figure}[t]
\begin{small}
\hrule
$$\begin{array}{r@{}l}
\SupportGuess  =  \{ &\mathit{variable}(X) \leftarrow  \mathit{var}(R,X),\\ 
&\mathit{suppSubst}(X,C) \vee \mathit{nsuppSubst}(X,C) \leftarrow \mathit{variable}(X), \mathit{dom}(C),\\
&\mathit{saturate} \leftarrow \mathit{suppSubst}(X,C_{1}), \mathit{suppSubst}(X,C_{2}), C_1 \neq C_2,\\ 
&\mathit{saturate} \leftarrow \mathit{unassigned}(X),\\
&\mathit{unass}(X,C)    \leftarrow \mathit{first}(C), \mathit{nsuppSubst}(X,C),\\ 
&\mathit{unass}(X,C_2)   \leftarrow \mathit{succ}(C_1,C_2), \mathit{unass}(X,C_1), \mathit{nsuppSubst}(X,C_2),\\ 
&\mathit{unassigned}(X) \leftarrow \mathit{last}(C), \mathit{unass}(X,C)\ \}\mbox{.} \\[.3em]
\SupportCheck  =  \{ &
\mathit{unfounded} \leftarrow \mathit{unsupp}(R), \mathit{lastR}(R),\\
&\mathit{unsupp}(R)   \leftarrow \mathit{firstR}(R), \mathit{unsuppRule}(R),\\
&\mathit{unsupp}(R_{2})  \leftarrow \mathit{succR}(R_{1},R_{2}), \mathit{unsupp}(R_{1}), \mathit{unsuppRule}(R_{2}),\\
& \mathit{saturate} \leftarrow \mathit{unfounded},\\
 &\mathit{suppSubst}(X,C)  \leftarrow \mathit{variable}(X),\mathit{dom}(C), \mathit{saturate},\\
&\mathit{nsuppSubst}(X,C) \leftarrow \mathit{variable}(X),\mathit{dom}(C), \mathit{saturate}\ \}\cup\\
\{&\mathit{unsuppRule}(R) \leftarrow c_{i}(R) \mid i \in \{1,\ldots, 5\}\}\mbox{.}
\end{array}$$
\hrule
\end{small}
\caption{Modules $\SupportGuess$ and  $\SupportCheck$\label{fig:SUPPORT}.}
\label{fig:support}
\end{figure}

\begin{definition}\label{def:unfounded}
By $\Support$ we understand the program $\SupportGuess \cup \SupportCheck \cup \SupportAux$, where
$\SupportGuess$ and $\SupportCheck$ are given in Figure~\ref{fig:support}, 
and $\SupportAux$ defines the auxiliary predicates $\mathit{succ}(\cdot,\cdot)$, $\mathit{succR}(\cdot,\cdot)$, $\mathit{first}(\cdot)$, $\mathit{last}(\cdot)$, $\mathit{firstR}(\cdot)$, $\mathit{lastR}(\cdot)$, and  
$c_{1}(\cdot)\commadots c_{5}(\cdot)$.
\end{definition}

The intuition behind this definition is as follows.
Consider a program $P$, some set $J$ of ground literals, encoded via $\mathit{inLoop}(\cdot)$, and an  interpretation $I$.
Module $\SupportGuess$ non-deterministically guesses a binary  relation $\mathit{suppSubst}(\cdot,\cdot)$ 
between the variables and the constant symbols in $P$. 
In case this relation is  not a function, 
$\SupportGuess$ establishes $\mathit{saturate}$. 
Module $\SupportCheck$, in turn, encodes whether, for each substitution $\subst$ and each
rule $r \in P$, some of the  conditions from the definition of $J$ being externally supported by $P$ is violated.
In fact, $\mathit{unfounded}$ is derived if some of these conditions is violated.
Moreover, $\mathit{saturate}$ holds if $\mathit{unfounded}$ holds, and $\SupportCheck$ saturates the
relation defined by predicate $\mathit{suppSubst}(\cdot,\cdot)$ if $\mathit{saturate}$ holds. 
Module $\SupportAux$ (omitted for space reasons) defines
 $\mathit{succ}(\cdot,\cdot)$ and $\mathit{succR}(\cdot,\cdot)$, which express the immediate successor relation, based on $\preceq$, for the constant symbols and rules in $P$, respectively, as well as
the predicates $\mathit{first}(\cdot)$, $\mathit{firstR}(\cdot)$, $\mathit{last}(\cdot)$, and $\mathit{lastR}(\cdot)$, which mark the first and the last elements 
in the order
defined by $\mathit{succ}(\cdot,\cdot)$ and $\mathit{succR}(\cdot,\cdot)$, respectively.
Moreover, the module $\SupportAux$ defines predicates $c_{1}(\cdot)\commadots c_{5}(\cdot)$, expressing failure of one of the conditions for $J$ being externally supported by $P$ \wrt\ $I$.

The rough idea behind the encoded saturation technique is to search, via 
$\SupportGuess$, for  counterexample substitutions that
witness that the  set $J$ of ground literals 
is \emph{not} unfounded. For such a substitution, neither  
$\mathit{saturate}$ nor $\mathit{unfounded}$ can become true which implies
that no answer set can contain $\mathit{unfounded}$ due to the saturation of $\mathit{suppSubst}(\cdot,\cdot)$ and the minimality of answer sets.
 
\begin{theorem}\label{thm:support}
Consider a program $P$, an interpretation $I$, and a set $J$ of ground literals. Then,
$J$ is unfounded by $P$ \wrt\ $I$ iff
the unique answer set of $\Support\ \cup\ \INPUT\ \cup\ \{\mathit{inLoop}(x') \mid x \in J\}$ contains the literal $\mathit{unfounded}$.
\end{theorem}

Given the above defined program modules, 
we arrive at the  uniform  encoding of the overall program $\Gamma$.

\begin{definition}
Let $\Unsat$, $\Loop$, and $\Support$ be the programs from Definitions~\ref{def:unsat}, \ref{def:loop}, and \ref{def:unfounded}, respectively. Then,  
$\Gamma = \Unsat\ \cup\ \Loop\ \cup\ \Support\ \cup\ \Cons$, {where}
\begin{center}
$\begin{array}{r@{}l}
 \Cons  =  
 \{ & \mathit{notAnswerSet} \leftarrow \mathit{unsatisfied},
\quad \mathit{notAnswerSet} \leftarrow \mathit{isLoop}, \mathit{unfounded}, 
\\
&\leftarrow \naf\ \mathit{notAnswerSet} \}\mbox{.} 
\end{array}$
\end{center}
\end{definition}

Module $\Cons$ encodes that each answer set of $\Gamma$ witnesses either $I \not\models P$ or that some loop $L \subseteq I$ of $P$ is unfounded by $P$ \wrt\ $I$. 

We finally obtain our main result, which follows essentially from the semantics of module $\Cons$ and
Theorems~\ref{thm:unsat}, \ref{thm:loop}, and \ref{thm:support}.

\begin{theorem}\label{th:main}
Given a program $P$ and an interpretation $I$, $\Pi=\Gamma \cup \INPUT$ satisfies the following properties:
\begin{enumerate}[\rm(i)]

\item  $\Pi$ has no answer set iff $I$ is an answer set of $P$.

\item $I$ is not an answer set of $P$ iff, for each answer set $S$ of  $\Pi$, $\{\mathit{unsatisfied},\mathit{unfounded}\}\cap S\neq\emptyset$.

\item $I \not\models P$ iff $\mathit{unsatisfied}\in S$, for some answer set $S$ of $\Pi$.
Moreover, for each rule  $r \in P$  with $I \not\models r\subst$, for some substitution $\subst$ over the Herbrand universe of $P$, there is some answer set $S$ of $\Pi$ such that (a)~$\{\mathit{unsatisfied},\mathit{guessRule}(r')\}\subseteq S$ 
and (b)~$\mathit{subst}(x',c)\in S$ iff $\subst(x) = c$.

\item A loop $L \subseteq I$ is unfounded by $P$ \wrt\ $I$ iff some answer set $S$ of $\Pi$ 
contains both $\mathit{isLoop}$ and $\mathit{unfounded}$, and
$L = \{x \mid \mathit{inLoop}(x') \in S\}$.
\end{enumerate}
\end{theorem}

\section{Applying the debugging approach}

In this section, we  first describe a simple scenario with different debugging tasks and 
show how the meta-program defined in the previous section can be used to solve them.
Afterwards, we discuss some pragmatic aspects relevant for realising a prospective user-friendly debugging system based on our approach.

\subsection{A simple debugging scenario}\label{sec:peanuts}

We assume that students have to encode the assignments of  papers to members of a program committee (PC) based on some bidding information in terms of ASP.
We consider three cases, each of them illustrates a different kind of debugging problem.
In the first case, an answer set is expected but the program is inconsistent. In the
second case, multiple answer sets are expected but the program yields only one answer set.
In the third case, it is expected that  a program is inconsistent, but it actually yields some answer set.
We illustrate that, in all cases, our approach gives valuable hints
how to debug the program in an iterative way.

Assume that $\mathit{pc}(X)$ means that $X$ is a member of the PC,  $\mathit{paper}(X)$ means that $X$ is a paper,
and $\mathit{bids}(X,Y,Z)$ means that PC member $X$ bids on paper $Y$ with value $Z$, where $Z$ is a natural number 
ranging from $0$ to $3$
expressing a degree
of preference for that paper.

To start with, Lucy wants to express that the default bid for a paper is $1$. That is, if a PC member does not bid on a paper, then 
it is assumed that the PC member bids $1$ on that paper per default.
Lucy's first attempt looks as follows:
\begin{center}
$\begin{array}{r@{}l}
L_{1} = \{& \mathit{pc}(m_1), \mathit{pc}(m_2), \mathit{paper}(p_1),
            \mathit{bid}(m_1,p_1,2), \mathit{bid}(m_2,p_1,3),\\
           & \mathit{some\_bid}(M,P) \leftarrow \mathit{bid}(M,P,X),\\
           & \mathit{bid}(M,P,1) \leftarrow \naf\ \mathit{some\_bid}(M,P),
            \mathit{pc}(M), \mathit{paper}(P)\}\mbox{.}
\end{array}$
\end{center}

Lucy's intention is that $\mathit{some\_bid}(M,P)$ is true if PC member $M$ bids on paper $P$, and 
$\mathit{bid}(M,P,1)$ is true if there is no evidence that PC member $M$ has bid on that paper.
Indeed, the unique answer set of $L_{1}$ is 
$$
\begin{array}{r@{}l}
S_{1} = 
\{& \mathit{pc}(m_1), \mathit{pc}(m_2), \mathit{paper}(p_1), \mathit{bid}(m_1,p_1,2),
\mathit{bid}(m_2,p_1,3), 
 \\&
\mathit{some\_bid}(m_1,p_1), \mathit{some\_bid}(m_2,p_1)  \}\mbox{.} 
\end{array}
$$
The answer set $S_{1}$ is indeed as expected:
We have that each PC member bids on some paper in $L_{1}$ and the 
last rule is inactive. Lucy's next step is to delete the fact $\mathit{bid}(m_2,p_1,3)$ from $L_{1}$---let us denote the 
resulting program by $L_{2}$.  Lucy expects that the answer set of $L_{2}$ contains $\mathit{bid}(m_2,p_1,1)$.
However, it turns out that $L_{2}$ yields no answer set at all!

To find out what went wrong, Lucy defines her expected answer set  as 
$$E_{1} = (S_{1} \cup \{\mathit{bid}(m_2,p_1,1)\}) \setminus 
            \{ \mathit{some\_bid}(m_2,p_1), \mathit{bid}(m_2,p_1,3)\}$$
and inspects the answer sets of $\Gamma \cup \reif{L_{2},E_{1}}$. 
It turns out that one answer set contains the facts $\mathit{unsatisfied}$ and $\mathit{guessRule}(r_{1}')$, where $r_{1}'$ is the label for the rule 
$$r_{1} = \mathit{some\_bid}(M,P) \leftarrow \mathit{bid}(M,P,X)\mbox{.}$$
Hence, $r_{1}$ is not satisfied by $E_{1}$:
$\mathit{bid}(m_2,p_1,1)$ is in $E_{1}$ and thus satisfies the body of $r_{1}$, but the head of $r_{1}$ is not
satisfied since $E_{1}$ does not contain $\mathit{some\_bid}(m_2,p_1)$.

Now that Lucy sees that $L_{2}$'s answer set has to contain $\mathit{some\_bid}(m_2,p_1)$, she defines
$E_{2}$ as $E_{1}$ plus the fact $\mathit{some\_bid}(m_2,p_1)$.
The answer sets of  $\Gamma \cup \reif{L_{2},E_{2}}$ reveal that $E_{2}$ is not an answer set of $L_{2}$ because
the singleton  loop $\mathit{bid}(m_2,p_1,1)$ is contained in $E_{2}$ but it is unfounded
by $L_{2}$ \wrt\ $E_{2}$.
The reason is clear:  the only rule that could support $\mathit{bid}(m_2,p_1,1)$ is 
$$r_{2} = \mathit{bid}(M,P,1) \leftarrow \naf\ \mathit{some\_bid}(M,P), \mathit{pc}(M),  \mathit{paper}(P)\mbox{.}$$
However, $r_{2}$ is
blocked since $E_{2}$ contains $\mathit{some\_bid}(m_2,p_1)$.

Lucy concludes that, to make $r_{2}$ work as expected, $\mathit{some\_bid}(m_2,p_1)$ must not be contained in the answer set.
To achieve this, Lucy changes $r_{1}$, the only rule with predicate $\mathit{some\_bid}$ in the head, into
$$\mathit{some\_bid}(M,P) \leftarrow \mathit{bid}(M,P,X), X \neq 1\mbox{.}$$
The resulting program works as expected and contains $\mathit{bid}(m_2,p_1,1)$ in its answer set.

The next student who is faced with a mystery is Linus.
He tried to formalise that each paper is non-deterministically assigned to at least one member of the PC.
His program  looks as follows:
\begin{center}
$\begin{array}{r@{}l}
P_{1} = \{ & \mathit{pc}(m_1), \mathit{pc}(m_2), \mathit{paper}(p_1), \mathit{paper}(p_2), \mathit{bid}(m_1,p_1,2),\\
      &         \mathit{bid}(m_1,p_2,3),\mathit{bid}(m_2,p_1,1), \mathit{bid}(m_2,p_2,1),
              \\
         &   \mathit{assigned}(P,M) \vee \neg \mathit{assigned}(P,M) \leftarrow \mathit{paper}(P),
            \mathit{pc}(M),\\
&\leftarrow \mathit{paper}(P), \mathit{pc}(M), \naf\ \mathit{assigned}(P,M) \}\mbox{.}
\end{array}$
\end{center}

Linus expects that the disjunctive rule realises the non-deterministic guess, and then the constraint prunes away all answer set candidates
where a paper is not assigned to some PC member.
Now, poor Linus is  desperate since the non-deterministic guess seems not to work correctly; the only answer set  of $P_{1}$ is
\begin{center}
$\begin{array}{r@{}l}
S_{3} = \{ & 
 \mathit{paper}(p_1), \mathit{paper}(p_2), \mathit{pc}(m_1), \mathit{pc}(m_2), \mathit{bid}(m_1,p_1,2), \mathit{bid}(m_1,p_2,3),  \\& 
 \mathit{bid}(m_2,p_1,1), \mathit{bid}(m_2,p_2,1), \mathit{assigned}(p_1,m_1),  
  \mathit{assigned}(p_1,m_2), \\& 
  \mathit{assigned}(p_2,m_1), 
 \mathit{assigned}(p_2,m_2) \},
\end{array}$
\end{center}
although Linus expected one answer set for each possible assignment.
In particular, Linus expected 
$$E_{3} = (S_{3} \cup \{\neg\mathit{assigned}(p_1,m_2)\}) \setminus \{\mathit{assigned}(p_1,m_2)\}$$
to be an answer set as well. 
Hence, Linus inspects the answer sets of $\Gamma \cup \reif{P_{1},E_{3}}$ and learns that the constraint in $P_{1}$
is not satisfied by $E_{3}$. In particular, it is the substitution that maps the variable $P$ to $p_1$ and $M$ to $m_2$ that is responsible for 
the unsatisfied constraint, which can be
seen from the $\mathit{subst}(\cdot)$ atoms in each answer set that contains $\mathit{unsatisfied}$. 

Having this information, Linus observes that the constraint in its current form is unsatisfied if  some paper is not assigned to \emph{each} PC member.
However, he intended it to be unsatisfied only when a paper is assigned to \emph{no} PC member.
Hence, he replaces the constraint
by the two rules 
$$\leftarrow \mathit{paper}(P), \mathit{pc}(M), \naf\ \mathit{at\_least\_one}(P)
\quad\mbox{and}\quad
\mathit{at\_least\_one}(P) \leftarrow \mathit{assigned}(P,M)\mbox{.}$$
The resulting program yields the nine expected answer sets.

Meanwhile, Peppermint Patty encounters a strange problem.
Her task was to  write a program that expresses the following issue:
If a PC member $M$ bids 0 on some paper $P$, then this means that there is  a conflict of interest \wrt\ $M$ and $P$.
In any case, there is a conflict of interest if  $M$ (co-)authored $P$. A PC member can only be assigned to some paper if
there is no conflict of interest \wrt\ that PC member and that paper.
This is Peppermint Patty's solution:
\begin{center}
$\begin{array}{r@{}l}
Q_{1} = \{ & \mathit{pc}(m_1),  \mathit{paper}(p_1), \mathit{bid}(m_1,p_1,2), \mathit{assigned}(p_1,m_1), \mathit{author}(p_1,m_1), \\&
\mathit{conflict\_of\_interest}(M,P) \leftarrow \mathit{bid}(M,P,0), \\&
\mathit{conflict\_of\_interest}(M,P)  \leftarrow \mathit{pc}(M), \mathit{paper}(P), \mathit{author}(M,P),\\&
\mathit{bid}(M,P,0) \leftarrow \mathit{pc}(M), \mathit{paper}(P), \mathit{conflict\_of\_interest}(M,P),\\&
\leftarrow \mathit{assigned}(P,M), \mathit{bid}(M,P,0) \}\mbox{.}
\end{array}$
\end{center}
The facts in $Q_{1}$ should model a scenario where a PC member authored a paper and is assigned to that paper. 
According to the specification from above, this should not  be  allowed. Since Patty is convinced that her encoding is 
correct, she expects that $Q_{1}$ has no answer sets. But $Q_{1}$ has
the unique answer set 
$$S_{4}=\{\mathit{assigned}(p_1,m_1),\mathit{pc}(m_1),
\mathit{paper}(p_1),
\mathit{author}(p_1,m_1),\mathit{bid}(m_1,p_1,2) \}\mbox{.}$$
What Peppermint Patty finds puzzling is that $S_{4}$ does not contain any atoms signalling a conflict of interest.
Hence, she decides to analyse why 
$$E_{4} = S_{4} \cup \{\mathit{conflict\_of\_interest}(m_1,p_1),\mathit{bid}(m_1,p_1,0)\}$$
is not an answer set of $Q_1$. 
If $Q_{1}$ was correct, then the only reason why $E_{4}$ is not an answer set of $Q_{1}$ would be
that the (only) constraint in $Q_{1}$ is unsatisfied. 

As expected, some answer sets of $\Gamma \cup \reif{Q_{1},E_{4}}$ contain 
 $\mathit{unsatisfied}$ and $\mathit{guessRule}(r')$, where $r'$ is the label of the constraint in $Q_{1}$. However, some answer sets contain the atom $\mathit{unfounded}$ 
as well---a surprising observation. Patty learns, by inspecting the $\mathit{inLoop}(\cdot)$ atoms in the respective answer set, that $E_{4}$ contains the loop 
$$\{\mathit{conflict\_of\_interest}(m_1,p_1),\mathit{bid}(m_1,p_1,0)\}$$
which is unfounded by $Q_{1}$ \wrt\ $E_{4}$:  $\mathit{bid}(m_1,p_1,0)$ seems to be justified only by the literal 
$\mathit{conflict\_of\_interest}(m_1,p_1)$ and vice versa. 
This should not be the case since $Q_{1}$ contains the rule 
$$\mathit{conflict\_of\_interest}(M,P)  \leftarrow \mathit{pc}(M), \mathit{paper}(P), \mathit{author}(M,P)$$
that should support $\mathit{conflict\_of\_interest}(m_1,p_1)$ because all the facts $\mathit{pc}(m_1)$, $\mathit{paper}(p_1)$, and $\mathit{author}(m_1,p_1)$ should
be contained in $Q_{1}$. Now, the error is obvious: $Q_{4}$ does not contain the fact $\mathit{author}(m_1,p_1)$ but $\mathit{author}(p_1,m_1)$---the order of the arguments was wrong.
After Peppermint Patty fixed that bug, her program is correct.  

\subsection{Some pragmatic issues and future prospects}

For a debugging system of practical value, certain pragmatic aspects have to be taken into account which we briefly sketch in what follows.
To start with, our encodings can be seen as a ``golden design''---tailored towards
clarity and readability---which leaves room for optimisations. 
Related to this issue, solver features like limiting the number of computed answer sets or query answering 
are needed to avoid unnecessary computation and to limit the amount of information presented to the user. 

Our debugging approach requires information about the intended semantics in form of
the interpretation representing a desired answer set.
Typically, answer sets of programs encoding real-world problems tend to  be large
which makes it quite cumbersome to manually create interpretations  from scratch.
It is therefore vital to have convenient means
for obtaining  an intended answer set in the first place.
For this purpose, we envisage a tool-box for managing  interpretations that allows
for their manipulation and storage. 
In such a setting, answer sets of previous versions of the debugged program could be a valuable source of  interpretations which are then tailored towards an intended answer set of the current version.
In addition to manual adaptations, partial evaluation of the program could significantly accelerate the creation of interpretations. 
We plan to further investigate these issues and
aim at incorporating our debugging technique, along with an interpretation management system as outlined, in an integrated development environment (IDE).
Here, an important issue is to achieve a suitable user interface for highlighting the identified unsatisfied rules and unfounded loops  in the source code and for visualising the involved variable substitutions.

\section{Related work}

Besides the debugging approach by \citeN{gebser08}, as already discussed earlier, other related approaches on debugging include the work of \citeANP{pontelli09} \citeyear{pontelli09} on \emph{justifications} for non-ground answer-set programs that can be seen as a complementary approach to ours.
Their goal is to explain the truth values of literals with respect to a given actual answer set of a program.
Explanations are provided in terms of \emph{justifications} which are labelled graphs
whose nodes are truth assignments of possibly default-negated ground atoms.
The edges represent positive and negative support relations between these truth assignments such that every path ends in an assignment which is either assumed or known to hold.
The authors have also introduced justifications for partial answer sets that emerge during the solving process (online justifications), being represented by three-valued interpretations.

The question why atoms are contained or are not contained in an answer set has also been raised by \citeANP{brain05} \citeyear{brain05} who provide algorithms for recursively computing explanations in terms of satisfied supporting rules.
Note that these problems can in principle also be handled by our approach, as illustrated in Section~\ref{sec:peanuts}.
Indeed, consider some program $P$ with answer set $I$ and suppose we want to know why a certain set $L$ of literals is contained in $I$.
Using our approach, explanations why $I\setminus L$ is not an answer set of $P$ will reveal rules which are unsatisfied under $I\setminus L$ but which support literals in $L$ under $I$.
Likewise, we can answer the question why expected atoms are missing in an answer set. 

\citeN{syrjaenen06} aims at finding explanations why some propositional program has no answer sets. 
His approach is based on finding minimal sets of constraints such that their removal yields consistency.
Hereby, it is assumed that a program does not involve circular dependencies
between literals through an odd number of negations which might also cause inconsistency.
Finding reasons for program inconsistency can be  handled by our approach when an intended answer set is known, as illustrated by program $L_2$ in Section~\ref{sec:peanuts}.
Otherwise, an interpretation can be chosen from the answer sets resulting from temporarily removing all constraints from the considered program (providing this yields consistency). 

\citeN{brain07} rewrite a program using some additional control
atoms, called \emph{tags}, that allow, e.g., for switching individual rules on or off and for analysing the resulting answer sets.
Debugging requests in this approach can be posed by adding further rules that can employ tags as well.
One such extension allows also for detecting atoms in unfounded loops.
However, as opposed to our current approach, the individual loops themselves are not identified.

\citeN{CaballeroGS08} developed a declarative debugging approach for datalog
using a classification of error explanations similar to the one by \citeN{gebser08} and our current work.
Their approach is tailored towards query answering  and,
in contrast to our approach, the language is restricted to stratified datalog. 
However, \citeANP{CaballeroGS08} provide an implementation that is based on computing
a graph that reflects the execution of a query.

\citeN{WittocxVD09} show how a calculus can be used for
debugging  first-order theories with inductive definitions in the context of model expansion problems, \iec problems of finding models of a given theory that expand some given interpretation. 
The idea is to trace the proof of inconsistency of such an unsatisfiable model expansion problem.
The authors provide a system that allows for interactively exploring the proof tree.

Besides the mentioned approaches which rely on the semantical behaviour of programs, \citeANP{mirek07} \citeyear{mirek07} use a translation from logic-program rules to natural language in order to detect program errors more easily.
This seems to be a potentially useful feature for an IDE as well, especially for novice and non-expert ASP programmers.

\section{Conclusion}

Our approach for declaratively debugging non-ground an\-swer-set programs
aims at providing intuitive explanations why a given interpretation fails to be an
answer set of the program in development.
To answer this question, we localise, on the one hand, unsatisfied rules
and, on the other hand, loops of the program that are unfounded with respect
to the given interpretation.
As underlying technique, we use a sophisticated meta-programming method
that reflects the complexity of the considered debugging question
which resides on the second level of the polynomial hierarchy.

Typical errors in ASP may have quite different reasons and
many of them could be avoided rather easily in the first place, \egc 
by a compulsory declaration of predicates \cite{brain05},
forbidding uneven loops through negation~\cite{syrjaenen06}, 
introducing type checks, or  defining
program interfaces. 
We plan to realise these kinds of simple prophylactic techniques for our future IDE for ASP that will incorporate our current debugging approach.
In this context,  courses on logic programming at our institute shall provide a permanent testbed for our techniques.
Moreover, as part of an ongoing research project on methods and methodologies for developing answer-set programs~\cite{mmdasp}, we want to put research efforts into methodologies that avoid or minimise debugging needs right from the start.
As a next direct step regarding our  efforts towards debugging,
we plan to extend our approach to language features like aggregates, function symbols, and optimisation techniques such as minimise-statements or weak constraints.



\end{document}